\documentclass[amsmath,twocolumn,prb,superscriptaddress,longbibliography,notitlepage,nofootinbib]{revtex4-2}
\usepackage[colorlinks=true,allcolors=blue]{hyperref}
\usepackage{amsmath}
\usepackage{amssymb}
\usepackage{amsthm}
\usepackage{xcolor}
\usepackage{graphicx}
\usepackage{float}
\usepackage{subfigure}
\usepackage{braket}
\usepackage{hhline}

\begin{document}

\title{Gauge theory description of Rydberg atom arrays with a tunable blockade radius}

\author{Yanting Cheng}
    \affiliation{Institute of Theoretical Physics and Department of Physics, University of Science and Technology Beijing, Beijing 100083, China}
\author{Chengshu Li}
\email{lichengshu272@gmail.com}
    \affiliation{Institute for Advanced Study, Tsinghua University, Beijing 100084, China}
\date{\today}

\begin{abstract}
We discuss a Rydberg atom chain with a tunable blockade radius from the gauge theoretic perspective. When the blockade radius is one lattice spacing, this system can be formulated in terms of the PXP model, and there is a $\mathbb{Z}_2$ Ising phase transition known to be equivalent to a confinement--deconfinement transition in a gauge theory, the lattice Schwinger model. Further increasing the blockade radius, one can add a next-nearest neighbor (NNN) interaction into the PXP model. We discuss the interpretation of NNN interaction in terms of the gauge theory and how finite NNN interaction alters the deconfinement behavior and propose a corresponding experimental protocol. When the blockade radius reaches two lattice spacing, the model reduces to the PPXPP model. A novel gauge theory equivalent to the PPXPP model is formulated, and the phases in the two formulations are delineated. These results are readily explored experimentally in Rydberg quantum simulators. 
\end{abstract}

\maketitle

\section{Introduction}
Gauge theories play a pivotal role in our understanding of the fundamental interactions, culminating in the overarching standard model~\cite{Polyakov1987,Schwartz2013}. On a more emergent level, gauge theories also serve as an important tool for the understanding of various condensed matter systems, from quantum spin liquids~\cite{Savary2016,Zhou2017} to quantum Hall effects~\cite{Tong2016}. Among various developments of the monument of gauge theories, the idea of regulating them on a lattice --- known as the lattice gauge theory (LGT)~\cite{Kogut1979,Kogut1983} --- provides both conceptual progress~\cite{Huang2019,Zache2019,Kebri2021,chen2022,Jad1} and closer relation to condensed matter/quantum simulation~\cite{Byrnes2006,Zohar2012,Wiese2013,Hauke2013,Zoller2014,Zoller2016,Yang2020,Tarabunga2022,Halimeh2022,Jad4}.

Recently, programmable Rydberg atom arrays attract considerable research interest thanks to their relevance both as interesting many-body systems and as potential quantum computing platforms~\cite{Chandrasekharan1997,Saffman2016,Ebadi2022}. Focusing on the former, there are two major frontiers. The first is to explore the ground state quantum phases and the transitions thereof~\cite{Keesling2019,Ebadi2021}. Highlights here include a quantum spin liquid with the atoms located on the bonds of a kagome lattice~\cite{Semeghini2021,Verresen2021,Cheng2021,Tarabunga2022,Samajdar2022,Giudici2022,Rhine}. The second is to probe dynamical processes, a surprising discovery along this line being the quantum scar states~\cite{Bernien2017,Turner2018a,Turner2018b,Ho2019,Serbyn2021,Yao2022,Jad2,Jad3} that violate the eigenstate thermalization hypothesis. It turns out that the PXP model and its generalizations constitute a simple yet powerful modelling incorporating the all-important blockade effect in these systems. In this simplest case, the Rydberg blockade effect prescribes that two nearest-neighbor atoms can not be simultaneously excited to the Rydberg state, and the Hamiltonian reads
\begin{equation}
\hat{H}=\sum_i \hat{P}_{i-1}\hat{X}_{i}\hat{P}_{i+1}-m\sum_i\hat{Z}_i,\label{eq:pxp}
\end{equation}
where one uses spin up (down) to denote the Rydberg (ground) state, $\hat{P}_i=(1-\hat{Z}_i)/2$ is a projector to the ground state, and $\hat{X},\hat{Z}$ are the Pauli X, Z operators. The first term couples the ground state and the Rydberg state with a Rabi oscillation, and the second term originates from detuning.  Focusing on the ground state, there are two phases --- a disordered phase when $m\to-\infty$ where all the atoms are in the ground state, and a $\mathbb{Z}_2$ translation symmetry breaking phase as $m\to\infty$ where the atoms are alternatively excited to the Rydberg state. The phase transition in between is of the Ising universality~\cite{Slagle2021,Yao2022}.

A close interplay between the PXP models and the LGT becomes manifest thanks to joint efforts from the condensed matter/cold atom physics and the high energy physics communities~\cite{Yang2020,Tarabunga2022}. On one hand, the PXP models can be reformulated as LGTs, enabling application of gauge theory concepts to the understanding of the former. On the other hand, LGTs of interest can be recast into the PXP models, allowing experimental simulations via the Rydberg blockade. This is particularly interesting when one realizes that confinement, a phenomenon from gauge theory that is of great significance yet hard to directly measure, can be approached on a tabletop setting via this mapping~\cite{Cheng2022, mapping}. The two phases mentioned above, then, correspond to confined and deconfined phases in the LGT language.

A finer model including longer-range interactions between Rydberg atoms reads
\begin{equation}
\hat{H}=\sum_i \hat{X}_{i}-m\sum_i\hat{Z}_i+V_0\sum_{i<j}\frac{\hat{n}_i\hat{n}_j}{|r_i-r_j|^6},\label{eq:fine}
\end{equation}
where $\hat{n}_i=(\hat{Z}_i+1)/2$ is the Rydberg state number operator at site $i$ and one includes the $1/r^6$ van der Waals interaction. Throughout the paper we set the lattice spacing to be 1 so $r_i=i$. The typical energy scale $V_0$ can be translated to a typical length scale via $r_\mathrm{b}=(V_0/2)^{1/6}$, known as the Rydberg blockade radius. This length scale determines the range within which two atoms can not be simultaneously excited. Eq.~\eqref{eq:fine} reduces to the PXP model in Eq.~\eqref{eq:pxp} for $r_\mathrm{b}\simeq1$. Focusing on the case where $r_\mathrm{b}$ interpolates from 1 to 2, we keep only the next-nearest-neighbor (NNN) interaction and the model simplifies to
\begin{equation}
\hat{H}=\sum_i \hat{P}_{i-1}\hat{X}_{i}\hat{P}_{i+1}-m\sum_i\hat{Z}_i+V\sum_i\hat{n}_i \hat{n}_{i+2}.\label{eq:pxp_V}
\end{equation}
This PXP--$V$ model has been extensively studied in Refs.~\cite{Fendley2004,Lesanovsky2012,Giudici2019,Chepiga2019}; also see Refs.~\cite{Rader2019,Yu2022,Chepiga2021} for relevant discussions. Three major phases of disordered, $\mathbb{Z}_2$, and $\mathbb{Z}_3$ translation symmetry breaking, together with small floating phases have been identified. 

In this work, we generalize the PXP--LGT identification to the case with a tunable blockade radius, or the PXP--$V$ model. We focus on two particular parameter regimes. First, we ask how a finite $V$ modifies the confinement and deconfinement behavior of the system. While the original protocol introduced in Ref.~\cite{Cheng2022} does not seem to give consistent results for finite $V$, closer inspection shows that the underlying cause is kinetic instead of dynamic. We therefore propose a new protocol to probe the confinement--deconfinement transition that works for finite as well as vanishing $V$. Second, we zero in on the limit $V\to\infty$, $|m|\ll V$, where the model reduces to that of PPXPP, defined later in Eq.~\eqref{eq:ppxpp}. Given the different nature of constraint, we introduce a novel LGT that naturally incorporates the constraint and again boasts confined and deconfined phases. In both cases, we perform numerical simulations to support our analysis, and these proposals are readily carried out experimentally in Rydberg quantum simulators.

\section{Gauge theory perspective of the PXP--$V$ model}

\begin{figure}
    \centering
    \includegraphics[width=0.48\textwidth]{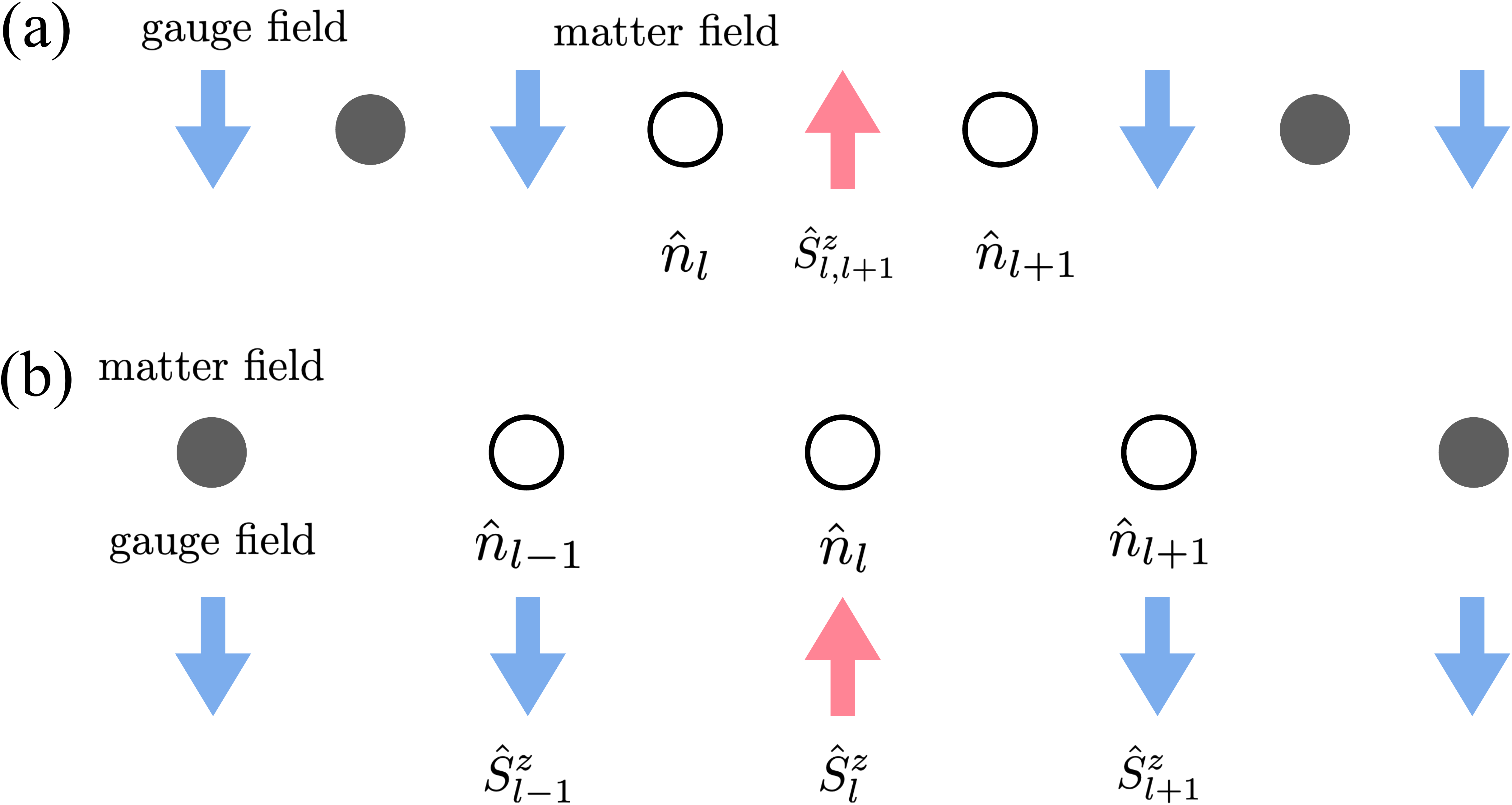}
    \caption{Schematic of the mapping between the PXP--$V$ model and the LGTs, where filled circles represent particles and empty circles represent holes. (a) When the NNN interaction is small, one obtains a modified lattice Schwinger model. (b) When the NNN interaction is strong enough, the Rydberg blockade radius is effectively 2 and one gets a novel LGT.}
    \label{fig:mapping}
\end{figure}

In this section we discuss the behavior of the system from the perspective of gauge theory with the Rydberg blockage radius interpolating from 1 to 2. As mentioned above, this can be described by the PXP--$V$ model, Eq.~\eqref{eq:pxp_V}. Before studying the effect of the NNN interaction, we first briefly review the PXP model, the lattice Schwinger model and the relation thereof. The PXP Hamiltonian dictates that two nearest-neighbor atoms can not be simultaneously excited to the Rydberg state, denoted by spin-up. In previous works, it is proven that the PXP model is equivalent to the U(1) LGT,
\begin{eqnarray}
    \hat{H}_{\mathrm{LGT}}=\sum_l\left[\hat{S}_{l,l+1}^+\hat{f}_l\hat{f}_{l+1}+h.c.+m\hat{n}_l^f\right]
\end{eqnarray}
in a particular subspace to be clarified shortly~\cite{Cheng2022, mapping}. Note that we now index the matter field fermions with $l$ and $\hat{n}_l^f=\hat{f}_l^\dagger\hat{f}_l$, and the gauge field spins are indexed with $(l,l+1)$, and $\hat{S}^z=\hat{Z}/2$. This U(1) LGT is nothing but the lattice Schwinger model. The action of the theta term in $1+1$D continuous quantum electrodynamics (QED), viz. the Schwinger model, reads
\begin{eqnarray}
    S_{\theta}=\int d^2x\frac{\theta}{2\pi}F_{01}.
\end{eqnarray}
Here $E=F_{01}$ is the $1+1$D electric field and the $\theta$ angle can be regarded as a background electric field, which affects the quantization of the electric field by $E=M+\frac{\theta}{2\pi}$ where $M$ is an integer. In the lattice model, the value of the electric field is $\pm1/2$ which means that the topological angle $\theta$ for this model is $\pi$. The lattice Schwinger model possesses local symmetries --- for each site, the transformation $\hat{f}_l\rightarrow e^{i\phi_l}\hat{f}_l$ with $\hat{S}^+_{l-1,l}\rightarrow e^{-i\phi_l}\hat{S}^+_{l-1,l}$ and $\hat{S}^+_{l,l+1}\rightarrow e^{-i\phi_l}\hat{S}^+_{l,l+1}$ leaves the Hamiltonian invariant. As a result, one can define a local conserved quantity at each site,
\begin{eqnarray}
    \hat{G}_l=\hat{S}_{l-1,l}^z+\hat{S}_{l,l+1}^z+\hat{n}^f_l,
\end{eqnarray}
which is the Gauss' law.

When $\hat{G}_l|\cdot\rangle=0$ for all $l$, the PXP model is equivalent to the lattice Schwinger model. This can be seen intuitively from the Fig.~\ref{fig:mapping}(a). For example, when the spin between site $l$ and site $l+1$ is spin-up, the matter fields in site $l$ and site $l+1$ are empty according to $\hat{G}_l|\cdot\rangle=0$. Then, because the matter field can not be annihilated from zero, it is no longer possible to flip the spins $\hat{S}_{l-1,l}$ and $\hat{S}_{l+1,l+2}$ to spin-up, which is the Rydberg blockade effect in the PXP model. 

\begin{figure*}
    \centering
    \includegraphics[width=\textwidth]{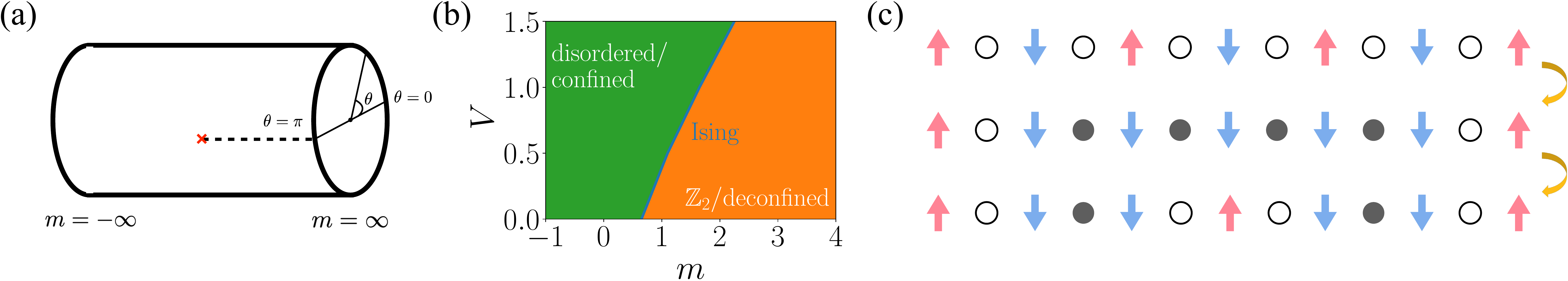}
    \caption{(a) The phase diagram of the lattice Schwinger model. $\theta$ is the topological angle which is defined by the quantization of the electric field (see main text for more information). (b) Part of the phase diagram of the PXP--$V$ model. (c) The protocol we propose to probe the confinement--deconfinement transition in the PXP--$V$ model.}
    \label{fig:phase}
\end{figure*}

In the previous work~\cite{Cheng2022}, one of the authors studied the PXP model from the gauge theory description by studying the phase diagram of the LGT which is shown in the Fig.~\ref{fig:phase}(a)~\cite{Coleman1976}. The dashed line represents the deconfined phase, and the complementary region is the confined phase. Roughly speaking, in the confined phase pulling away two particles costs an energy proportional to the distance in between and therefore not possible, while in the deconfined phase one is able to separate two particles with a finite energy cost. The protocol proposed to probe the confinement--deconfinement dichotomy is as follows. One first prepares a pair of fermions/holes by flipping one particular spin starting from the ground state. For example, when the ground state is that all the spins are spin-down and all the matter sites are filled, we flip a spin to spin-up as in Fig.~\ref{fig:mapping}(a) and then according to the Gauss' law we create a pair of holes next to the flipped spin. Then we use the correlation function
\begin{equation}
\mathcal{G}(r,t)=\sum_l\braket{\Psi(t)|(\hat{n}^f_l-\bar{n}^f_l)(\hat{n}^f_{l+r}-\bar{n}^f_{l+r})|\Psi(t)}\label{eq:G} 
\end{equation}
to characterize the distance between the two fermions/holes we added. Here $|\Psi(t)\rangle$ is the wavefunction after a spin is flipped and $\bar{n}_l$ denotes the mean particle number at site $l$ in the ground state before the spin is flipped. If the system is in the confined phase, the distance between these two fermions/holes is always 1; therefore the peak of the correlation function $\mathcal{G}(r,t)$ is always at $r=1$ for any $t$. If the system is in the deconfined phase, these two fermions/holes are able to get apart from each other; therefore the correlation function $\mathcal{G}(r,t)$ becomes distributed on the whole system as time goes by. As $m$ goes from $-\infty$ to $\infty$, Ref.~\cite{Cheng2022} finds that $\mathcal{G}(r,t)$ changes from being localized to being extended. One thus establishes an identification between the confinement--deconfinement transition and the $\mathbb{Z}_2$ transition, and gets an experimental access to the former.

\begin{figure*}
    \centering
    \includegraphics[width=\textwidth]{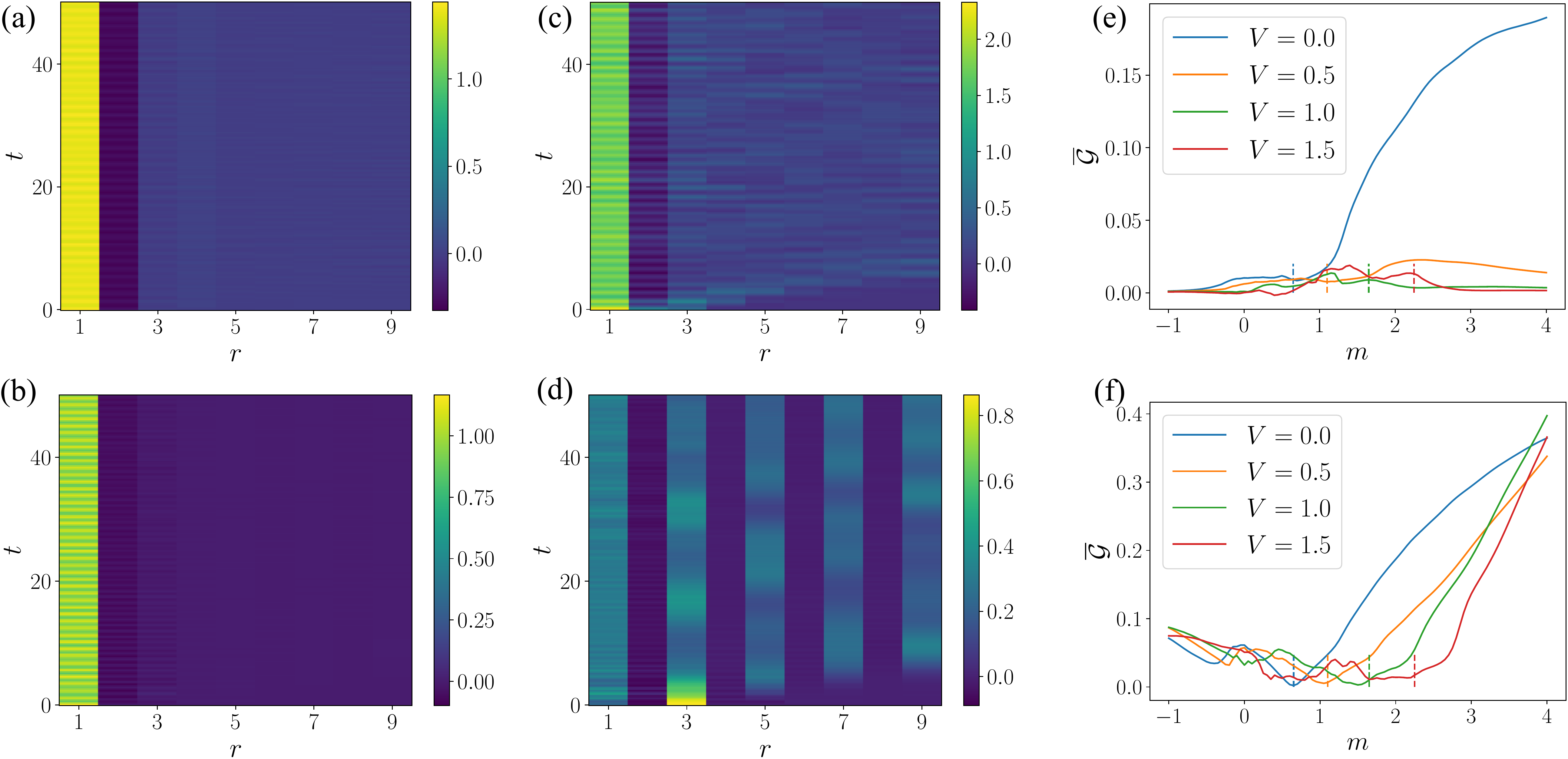}
    \caption{(a, b) The correlation function $\mathcal{G}(r,t)$ with the initial wavefunction prepared by the original protocol. (c, d) The same function obtained from the new protocol. We set $V=1.5, m=-1$ in (a, c) and $V=1.5, m=4$ in (b, d). (e, f) The averaged correlation function $\overline{\mathcal{G}}$ (see main text for definition), with the location of phase transition marked by the dashed lines, obtained from calculating the two-point correlation function. We take $N=18$ and periodic boundary conditions in all calculations.}
    \label{fig:pxp-res}
\end{figure*}

Now we wish to apply this scheme to the PXP--$V$ model, with the phase diagram in focus here shown in Fig.~\ref{fig:phase}(b). We first note that the gauge invariance is intact with the NNN term since it only involves $\hat{Z}$ operators. If one uses the aforementioned protocol literatim, however, numerical results do not seem to exhibit deconfinement behavior where they are expected to, see Fig.~\ref{fig:pxp-res}(a, b). Here we set $V=1.5$ and $m=-1, 4$ in the two subfigures. These two parameter choices lie in the disordered and ordered phases, respectively (cf. Fig.~\ref{fig:phase}(b)), and therefore the corresponding $\mathcal{G}(r,t)$ are expected to be localized and extended, contrary to the numerical results where both are localized. 

The key insight to solve this conundrum is the following identity
\begin{equation}
\begin{split}
&\hat{n}_l^f \hat{n}_{l+1}^f\\
={}&(-\hat{S}_{l-1,l}^z-\hat{S}_{l,l+1}^z)(-\hat{S}_{l,l+1}^z-\hat{S}_{l+1,l+2}^z)\\
={}&\hat{n}_{l-1,l}\hat{n}_{l+1,l+2}-\hat{n}_{l-1,l}-\hat{n}_{l,l+1}-\hat{n}_{l+1,l+2},\label{eq:id}
\end{split}
\end{equation}
where $\hat{n}^f$ denotes matter field fermion number operator and $\hat{n}$ denotes Rydberg atom number operator as usual. The second step follows from $\hat{n}_{l-1,l}\hat{n}_{l,l+1}=\hat{n}_{l,l+1}\hat{n}_{l+1,l+2}=0$. Using this, we have the Hamiltonian in the LGT language,
\begin{equation}
\begin{split}
    \hat{H}_{\mathrm{LGT-V}}=&\sum_l\Big[\hat{S}_{l,l+1}^+\hat{f}_l\hat{f}_{l+1}+h.c.+\Big(m-\frac{3V}{2}\Big)\hat{n}_l^f\\
    &+V\hat{n}_l^f \hat{n}_{l+1}^f\Big].
    \label{eq:lgt-v}
\end{split}
\end{equation}
Therefore, an NNN Rydberg interaction corresponds to a \emph{nearest-neighbor} interaction between the matter field particles, modulo terms that can be absorbed in the detuning. When one prepares a nearest-neighbor fermion/hole pair from flipping a spin, the new state has an interaction energy difference of $\pm V$. Energy conservation then forbids the pair to deconfine simply because particles far apart do not have the same interaction energy. Put another way, the disappearance of deconfinement behavior noted above is rooted in kinetic, instead of dynamic, rationale. By this we are referring to the difference between the local interaction energy of $V$ and the proportional-to-distance confining energy. 

The analysis above immediately suggests a cure to the original protocol, i.e., to create a non-nearest-neighbor fermion/hole pair when preparing the initial state. It turns out, as easily verified, that in all allowed cases the number of consecutive holes must be even. Hence a minimal choice is to create a next-next-nearest-neighbor fermion/hole pair. In the Rydberg/spin language, one can flip spins at sites $l,l+2,l+1$, in this order, see Fig.~\ref{fig:phase}(c). Note that spin flippings at neighboring sites do not commute.

We perform numerical simulations of this protocol, vis-\`a-vis the previous one for comparison. Two typical spacetime dependences of $\mathcal{G}(r,t)$ are shown in Fig.~\ref{fig:pxp-res}(c, d). For $m$ smaller than the critical value, $\mathcal{G}(r,t)$ remains localized as time evolves, while $\mathcal{G}(r,t)$ quickly spreads to the whole system range for large $m$. Note that for large enough systems we expect a light-cone-like behavior, and the reflections seen here are due to the finite sizes available to numeric simulations. To reduce this effect and distill a quantitative measure for the confinement--deconfinement transition, we introduce an averaged correlation function
\begin{equation}
\overline{\mathcal{G}}=\frac{\langle\mathcal{G}(r,t)\rangle_{r>4,t}}{\langle\mathcal{G}(r=1,t)\rangle_t}.
\end{equation}
That is, we average $\mathcal{G}(r,t)$ over $r>4$ and $t$ and normalize it against $\mathcal{G}(r=1,t)$ averaged over $t$. The averaged correlation functions from both protocols are shown in Fig.~\ref{fig:pxp-res}(e, f), where we also mark the phase transition points by vertical dashed lines, the locations of which pinpointed from the conventional two-point function. The inability of the original protocol to detect the deconfinement behavior with NNN interaction is clear from Fig.~\ref{fig:pxp-res}(e), where apart from the $V=0$ case $\overline{\mathcal{G}}$ remains small across the whole parameter range. With the new protocol, on the other hand, we see in Fig.~\ref{fig:pxp-res}(f) that the onset of deconfinement matches the phase transition point perfectly well. These results thus lend full support to our analysis.

Finally, we note that the Hamiltonian Eq.~\eqref{eq:lgt-v} provides an explanation of how the phase boundary tilts in Fig.~\ref{fig:phase}(b). The idea is that part of the detuning term can be combined with the NNN interaction term to give an equivalent nearest-neighbor matter field interaction, hence effectively reduces $m$ by an amount $\sim3V/2$. One therefore needs to go to a larger critical $m_c$ to arrive at the phase transition. The predicted $\Delta m_c\sim3\Delta V/2$ becomes exact for large $V$~\cite{Fendley2004}. This perspective complements the Rydberg/spin model interpretation that a finite positive $V$ costs an extra energy for the staggering spin configuration and therefore prefers the disordered state at a given $m$.

\section{Gauge theory description of the PPXPP model}
In increasing $V$, a new phase, namely the $\mathbb{Z}_3$ ordered phase comes out, while the $\mathbb{Z}_2$ phase is pushed to $m=\infty$. It turns out that the correlation function $\mathcal{G}(r,t)$ fails to capture this new piece of physics. To make progress, we focus on the $V\to\infty$ limit where the PXP--$V$ model reduces to the PPXPP model,
\begin{eqnarray}
    \hat{H}_{\mathrm{PPXPP}}=\sum_i\hat{P}_{i-2}\hat{P}_{i-1}\hat{X}_i\hat{P}_{i+1}\hat{P}_{i+2}-m\sum_i\hat{Z}_i.\label{eq:ppxpp}
\end{eqnarray}
That is, any two atoms within a range of 3 can not be excited to the Rydberg state simultaneously. The two limiting cases are easy to understand. When $m\to-\infty$, all the atoms are in the ground state and we are in the disordered phase. When $m\to\infty$, the translational symmetry is spontaneously broken to a $\mathbb{Z}_3$ phase. There also exists a small critical, incommensurate floating phase~\cite{Giudici2019}, but we will focus on the two major phases in this section.

\begin{figure*}
    \centering
    \includegraphics[width=\textwidth]{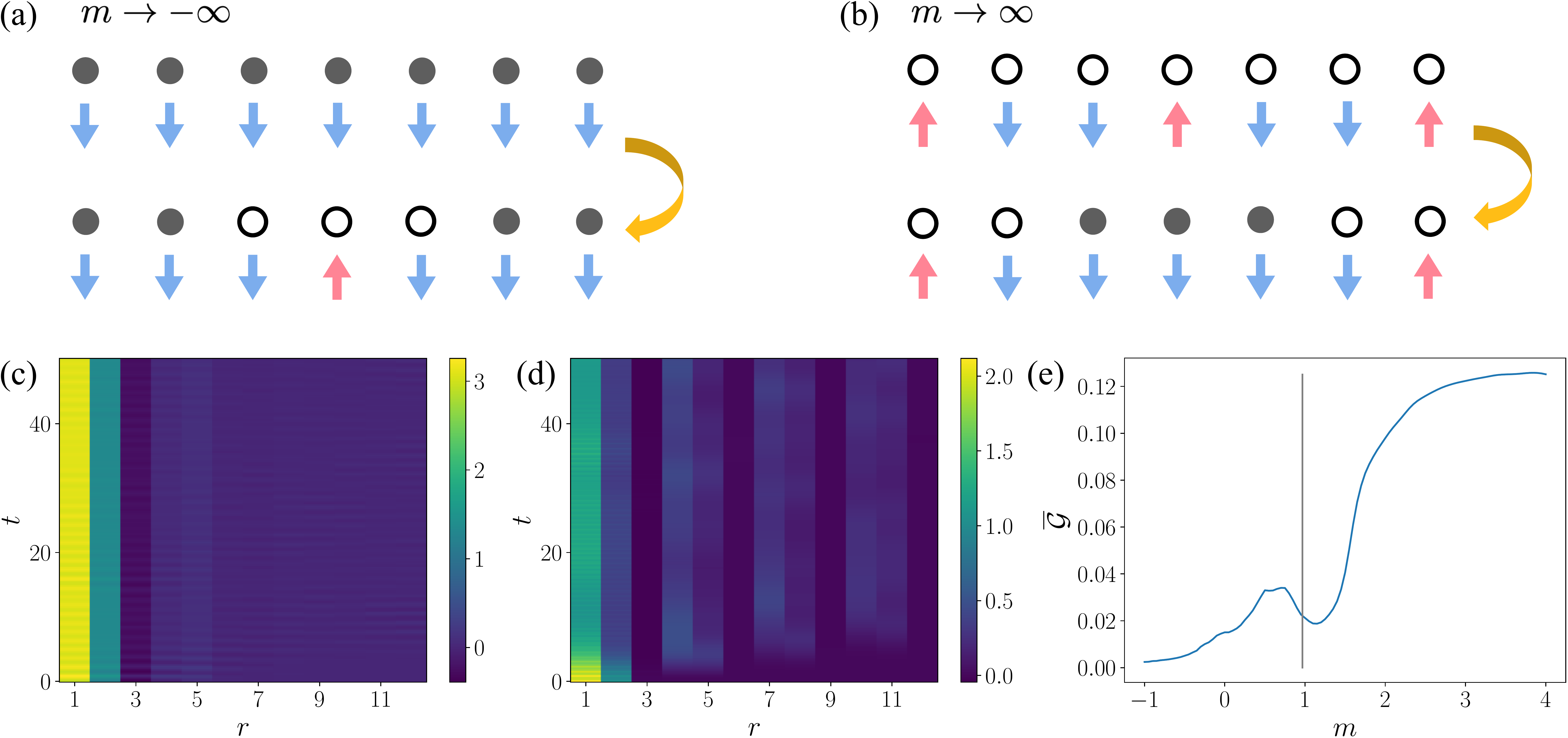}
    \caption{(a, b) Schematic for creation of particles/holes in the LGT corresponding to the PPXPP model in the $m\to\pm\infty$ limits. (c, d) The correlation function $\mathcal{G}(r,t)$. We set $m=-1$ in (c) and $m=4$ in (c). (e) The averaged correlation function $\overline{\mathcal{G}}$. The very narrow floating phase is plotted in grey, with its width approximately equal to the linewidth. We take $N=24$ and periodic boundary conditions in all calculations.}
    \label{fig:PPXPP}
\end{figure*}

We wish to similarly relate this model to an LGT. Apparently, the lattice Schwinger model fails to accommodate the constraint, but the same philosophy still works. We consider the following LGT,
\begin{eqnarray}
    \hat{H}_{\mathrm{LGT}}^{\mathrm{PPXPP}}=\sum_l\left[\hat{S}_l^+\hat{b}_{l-1}\hat{b}_l\hat{b}_{l+1}+h.c.+\frac{2}{3}m\hat{n}_l^b\right],
    \label{eq:LGT2}
\end{eqnarray}
which is equivalent to the PPXPP Hamiltonian in a particular subspace which will be clarified as follows. As shown in Fig.~\ref{fig:mapping}(b), now the matter fields and the gauge fields are living on the same lattice sites. For each site, there is a spin with a hard-core boson mode. The Hamiltonian Eq.~\eqref{eq:LGT2} possesses a local gauge symmetry that $\hat{b}_{l}\rightarrow\hat{b}_{l}e^{i\phi_l}$ together with $\hat{S}_{l-1}^+\rightarrow\hat{S}_{l-1}^+e^{-i\phi_l}$, $\hat{S}_{l}^+\rightarrow\hat{S}_{l}^+e^{-i\phi_l}$ and $\hat{S}_{l+1}^+\rightarrow\hat{S}_{l+1}^+e^{-i\phi_l}$ leaves the Hamiltonian invariant. The conserved quantity corresponding to this local symmetry is
\begin{eqnarray}
    \hat{G}_l^{\mathrm{PPXPP}}=\hat{S}_{i-1}^z+\hat{S}_{i}^z+\hat{S}_{i+1}^z+\hat{n}_i^b+\frac{1}{2}.
\end{eqnarray}
This lattice gauge model is equivalent to the PPXPP model when $\hat{G}_l^{\textrm{PPXPP}}|\cdot\rangle=0$ for all $l$.  Once a spin at site $i$ is flipped from spin-down to spin-up, three bosons are annihilated at site $i-1$, $i$ and $i+1$. Then the spins at site $i-2$, $i-1$, $i+1$ and $i+2$ can not be flipped to spin up any more, automatically satisfying the constraint in the PPXPP model. We note that the trilinear form of the matter field dictates use of bosons instead of fermions as prescribed by the superselection rule.

As mentioned before, there are two phases in the PPXPP model in the two limits of $m$. To manifest the corresponding LGT content, we again apply the scheme introduced in Ref.~\cite{Cheng2022}. In this case it turns out unnecessary to modify the original protocol, and flipping one single spin suffices. See Fig.~\ref{fig:PPXPP}(a, b) for an illustration of this process in the $m\to\pm\infty$ limits, where we add three holes/particles into the system simultaneously by flipping a spin. The numerical results of $\mathcal{G}(r,t)$, also defined with Eq.~\eqref{eq:G} albeit with redefined matter fields, are shown in Fig.~\ref{fig:PPXPP}(c, d). This correlation function also captures the distance between the holes/particles we add. Again, we see a clear distinction between confinement and deconfinement behavior as $m$ changes: for $m=-1$ (Fig.~\ref{fig:PPXPP}(c)), $\mathcal{G}(r,t)$ remains localized at $r=1$ for all $t$, while for $m=4$ (Fig.~\ref{fig:PPXPP}(d)), $\mathcal{G}(r,t)$ becomes distributed over the whole system. Similar to the PXP--$V$ case, we obtain a single indicator from $\mathcal{G}(r,t)$ by defining an averaged correlation function
\begin{equation}
\overline{\mathcal{G}}=\frac{\langle\mathcal{G}(r=11,t)\rangle_t}{\langle\mathcal{G}(r=1,t)\rangle_t},
\end{equation}
averaged only over time now which suffices for our purpose. Here $r=11$ is the maximal $r$ with non-trivial signals, see Fig.~\ref{fig:PPXPP}(d). Nevertheless, in experiments we expect any reasonably large $r$ to work as well. The correlation function $\overline{\mathcal{G}}$ is plotted in Fig.~\ref{fig:PPXPP}(e). A very narrow floating phase \cite{Giudici2019}, with its width approximately equal to that of the grey vertical line, separates the disordered and the $\mathbb{Z}_3$ phases. We see a consistent alignment between the onset of deconfinement behavior and the entrance to the ordered phase. This indicates that the disordered ($\mathbb{Z}_3$ ordered) phase can be identified with the (de)confined phase, a result similar to the PXP(--$V$) case.

\section{Conclusion}
In this work, based on previous works relating the PXP model describing a Rydberg atom chain with an LGT, we consider systems with a tunable Rydberg blockade radius. Such a system is conveniently modelled by the PXP--$V$ model, a subject extensively studied in the literature. We focus on two particular parameter regimes, one with small finite $V$ added to the PXP model and the other in the limit of the PPXPP model. In the first scenario we show how to modify the previously proposed protocol to incorporate the finite $V$ effect by elucidating the rationale of the latter. With the new protocol the confinement--deconfinement transition again becomes manifest. A gauge theoretic explanation of the phase boundary is also discussed, predicting a shifting of critical detuning of $\Delta m_c\sim 3\Delta V/2$. In the second scenario a novel LGT is formulated to automatically satisfy the PPXPP constraint. We establish an identification of the disordered ($\mathbb{Z}_3$ ordered) phase and the (de)confined phase. Both analyses are corroborated by numerical calculations and are expected to be of immediate experimental relevance.

While in Ref.~\cite{Fendley2004} the authors propose a direct chiral transition from the $\mathbb{Z}_3$ phase to the disordered phase in the parameter regime above the integrable Potts point, the authors of Ref.~\cite{Chepiga2019}, based on extensive numerical calculation, argue that an intermediate floating phase exists in between the two phases for large enough $V$. Calculations in Ref.~\cite{Giudici2019} suggest that this floating phase extends all the way to the $V\to\infty$ limit, the transition in focus here in the PPXPP model. A relation between this phase and the gauge theory hailed here is left for future investigation.

\section*{Acknowledgement}
We thank Hui Zhai and Shang Liu for helpful discussions. The project is supported by NSFC under Grant No. 12204034 and Grant No. 11874083, Fundamental Research Funds for the Central Universities (No.FRF-TP-22-101A1) and China Postdoctoral Science Foundation (Grant No. 2022M711868). C. L. is also supported by the International Postdoctoral Exchange Fellowship Program and the Shuimu Tsinghua Scholar Program. 
\bibliography{biblio}

\end{document}